\documentclass[twocolumn,prl,aps,nobalancelastpage]{revtex4}

\usepackage{graphicx}
\usepackage{dcolumn}
\usepackage{bm}

\begin{document}

\title{de Haas-van Alphen effect in single crystal MgB$_2$}

\author{E. A. Yelland and J. R. Cooper }
\affiliation{Interdisciplinary Research Centre in Superconductivity and
Department of Physics, University of Cambridge, Madingley Road, Cambridge
CB3 0HE, United Kingdom.}
\author{A. Carrington, N. E. Hussey and P. J. Meeson}
\affiliation{ H. H. Wills Physics Laboratory, University of Bristol, Tyndall Avenue, BS8 1TL, United Kingdom.}%
\author{S. Lee, A. Yamamoto and S. Tajima}
\affiliation{ Superconductivity Research Laboratory, International Superconducting Technology Center, Tokyo 135-0062, Japan.}%
\date{\today}

\begin{abstract}

We report observations of quantum oscillations in single crystals of the high temperature superconductor MgB$_2$. Three
de Haas-van Alphen frequencies are clearly resolved.  Comparison with band structure calculations strongly suggests
that two of these come from a single warped Fermi surface tube along the $c$ direction, and that the third arises from
cylindrical sections of an in-plane honeycomb network. The measured values of the effective mass range from $0.44-0.68
m_{e}$. By comparing these with band masses calculated recently by three groups, we find that the electron-phonon
coupling strength $\lambda$, is a factor $\sim 3$ larger for the $c$-axis tube orbits than for the in-plane network
orbit, in accord with recent microscopic calculations.
\end{abstract}

\maketitle

The  discovery of high temperature superconductivity in MgB$_2$ \cite{nagamatsu01} has led to great interest in this
hexagonal layered compound.  Early isotope effect work clearly showed that phonons are important \cite{budko01}, but
subsequent experiments and calculations by a number of groups have revealed some unusual features, such as the possible
existence of two distinct superconducting gaps \cite{bouquet01,chen01,manzano01,szabo01}, that may be associated with
two different bands \cite{liu01,shulga01}.

The bandstructure of MgB$_2$ has been calculated by a number of groups \cite{kortus01} and recently, photoemission
experiments \cite{uchiyama} have confirmed its qualitative features. Studies of the de Haas-van Alphen (dHvA) effect
provide crucial information about the electronic properties of metals and superconductors. Observation of quantum
oscillations allows the shape of the Fermi surface and the effective masses of carriers on individual Fermi sheets to
be found. Then, together with band-structure calculations, electron-phonon coupling constants can be obtained
\cite{shoenberg}. Here we report a detailed study of the dHvA effect in single crystals of MgB$_2$. Three dHvA
frequencies are clearly resolved in our data, and can be assigned to two distinct, orthogonal Fermi surfaces parallel
to the main symmetry axes of MgB$_2$ \cite{kortus01}.  The effective masses corresponding to these three frequencies
have been measured and compared  with recent calculations by three groups. The comparison shows clearly that the
electron phonon enhancement is large for the c-axis tube and much smaller for the in-plane electron-like tube.

The single crystals used in this work were grown in Tokyo by high pressure synthesis, as described in Ref.\
\cite{lee01}. Two crystals from the same batch were studied in parallel using sensitive piezo-resistive cantilevers to
measure the  torque ($\Gamma$) \cite{taillefer99,bergemannphd} -- crystal A at Cambridge using a $^4$He cryostat with a
15~T magnet and crystal B at Bristol using a $^3$He cryostat with an 18~T magnet. The dimensions of the crystals A and
B were $230\times 80\times 20\mu $m$^3$ and $230\times 200\times47\mu$m$^3$ respectively (the shortest side being along
the $c$-axis). Thermal contact between the crystals and nearby calibrated thermometers was ensured by immersion in
$^3$He or $^4$He liquids or a few mbars pressure of exchange gas. Both cryostats had a single-axis low-temperature
rotatable sample stage and angles were measured to a relative accuracy of at least 1$^\circ$. For crystal A, the
alignment was verified from the symmetry of high resolution, torque versus angle sweeps  near the $c$-axis at 15~T,
while for B a correction of 6$^\circ$ was applied to make the dHvA frequencies symmetrical about the crystal axes. The
data obtained in both sets of experiments were essentially identical.

\begin{figure}
\includegraphics[width=8.0cm,keepaspectratio=true]{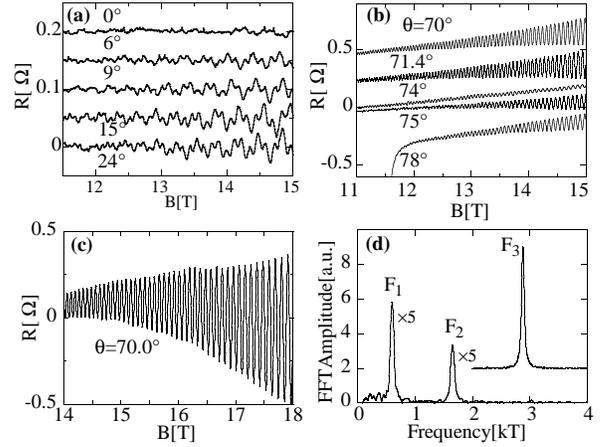}
 \caption{Panels a-c: dHvA signals at several angles, rotating from [001] to [100].
 Panel d: FFTs of the  data at $\theta=20^\circ$ ($F_1$ and $F_2$) and 70$^\circ$ ($F_3$).
 The top and bottom  panels show data for crystals A and B at 1.35~K and 0.36~K, respectively.}
 \label{wigglesfig}
\end{figure}

Fig.\ \ref{wigglesfig} (panels a-c) show torque versus magnetic field data \cite{footnote1} for both crystals as a
function of angle, as the crystals were rotated from $B\parallel$ [001] ($\theta=0^\circ$) to [100]
($\theta=90^\circ$). Initially, the orientation of the $c$-axis was determined from the platelet shape and the
anisotropy of the upper critical field $H_{c2}$. As expected, the torque signal goes to zero at the principal symmetry
directions ($\theta=0^\circ$ and 90$^\circ$) but is already sizable only a few degrees away. The signal was easily
resolved over a wide field range, giving well defined dHvA frequencies [see the fast Fourier transforms (FFTs) in Fig.\
\ref{wigglesfig} (panel d)]. Two frequencies ($F_1$ and $F_2$) were observed for field sweeps carried out within
$\sim$45$^\circ$ of the $c$-axis [Fig.\ \ref{wigglesfig}(a)], while only a single frequency $F_3$ was found for fields
nearer the plane [Fig.\ \ref{wigglesfig}(b) and (c)].

The dHvA signals were analyzed in the conventional way using the  Lifshitz-Kosevich (LK) expression
\cite{shoenberg,wasserman} for the oscillatory  magnetization of a 3D Fermi liquid
\begin{equation}
M_{osc} \propto B^{\frac{1}{2}} R_D R_T R_S \sin \left(\frac{2\pi F}{B}+\gamma\right) \label{lkexpression}
\end{equation}
where $F$ is the dHvA frequency [$F = (\hbar/2\pi e)A$,  $A$ is the extremal orbit area in $\mathbf{k}$-space];
$\gamma$ is the phase; $R_D$, $R_T$ and $R_S$ are the damping factors from impurity scattering, temperature~($T$) and
spin splitting respectively. $R_D$ = $\exp(-\pi m_B / eB\tau)$, where $m_B$ is the unenhanced or `bare' band mass
\cite{shoenberg,wasserman} and $\tau$ is the scattering time. $R_T = X/(\sinh X)$ where $X = (m^{*}/m_{e})(14.69/B)T$,
and $m^*$ is the quasi-particle effective mass, that is enhanced over $m_B$ by both electron-electron and
electron-phonon interactions, and $m_{e}$ is the free-electron mass. The spin splitting phase factor is given by $R_S =
\cos(n \pi g m_S/2m_e)$ where $g$ is the Land\'{e} $g$-factor for electrons in a given orbit, and $m_S/m_B$ is the
enhancement of the Pauli susceptibility from electron-electron interactions alone.

\begin{figure}
\includegraphics[width=7.5cm,keepaspectratio=true]{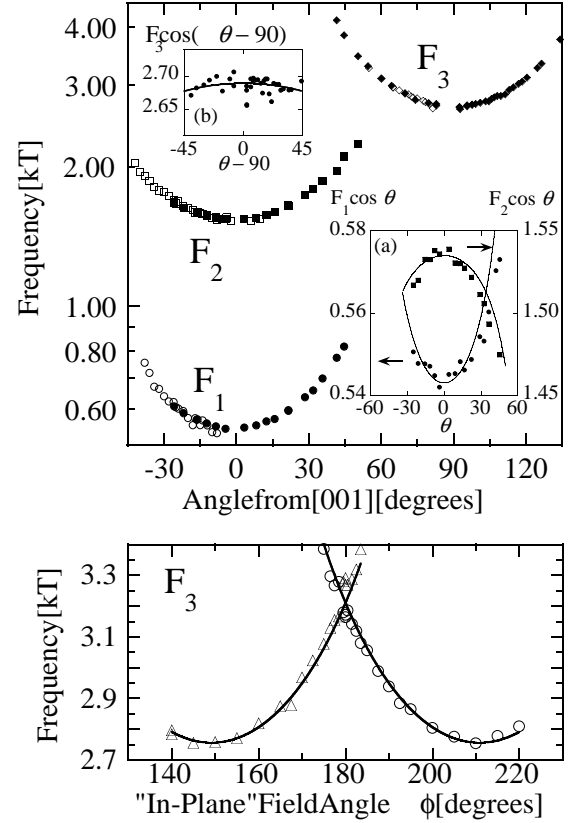}
 \caption{Upper Panel: Angular dependence of all three frequencies for a rotation between [001] and
 [100].
 The open/closed symbols are data for crystals A/B respectively. Inset (a) $F_1
 \cos\theta$ and $F_2 \cos\theta$, with a fit to Eq.\ (\protect\ref{yameq}).
 Inset (b) $F_3 \cos(\theta-90)$.
 Lower Panel: In-plane rotation plot for $F_3$, crystal A, with fits:
 2685/($|\cos\beta\sin\phi| +|\sin\beta\cos15\cos\phi| $) where
 $\beta$ = 60 or 120.}
 \label{rotationplot}
\end{figure}

Fig.\ \ref{rotationplot} (upper panel) shows the angular dependence of the three dHvA frequencies observed in both
crystals on rotating from $B\|$ [001] to [100]. There is excellent agreement between the two sets of data. The strong,
nearly cosine angular dependence of $F_1$ and $F_2$ suggests that they arise from cylindrical sections along the
$c$-axis. The small deviations from a strict $1/\cos\theta$ dependence are shown in inset (a) of Fig.\
\ref{rotationplot}. Assuming a simple cosine $c$-axis dispersion [i.e., $\varepsilon_k=\hbar^2(k_x^2+k_y^2)/2m
-2t_c\cos(ck_z)$], Yamaji \cite{yamaji89} derived a formula which accounts for the `magic-angle' magnetoresistance
maxima in quasi-2D organic superconductors. This gives the angular dependence of the two frequencies, $F_{\pm}$,
arising from extremal areas of a single warped cylinder as,
\begin{equation}
F_{\pm}(\theta) \cos\theta = (\hbar/2\pi e)[\pi k_f^2 \pm 4 \pi m t_c J_0(c k_f \tan\theta)]\label{yameq}
\end{equation}
In this expression, the first term is the mean frequency $(\hbar/2\pi e)\pi k_f^2=(F_1^0+F_2^0)/2$, the prefactor of
the second term $(\hbar/2\pi e)4 \pi m t_c$ is $(F_2^0-F_1^0)/2$ and $J_{0}$ is the Bessel function. Here $F_1^0$ and
$F_2^0$ represent $F_1$ and $F_2$ at $\theta=0$, and may be taken directly from the data, so this equation has no free
parameters.  The good agreement shown in inset (a) of Fig.\ \ref{rotationplot} implies that $F_1$ and $F_2$ are the two
extremal orbits of a single warped tube. As shown in inset (b), for the other orbit, $F_3 \cos(90-\theta)$ is
remarkably constant (to within $\pm$~1$\%$ over $\pm$ 40$^\circ$) implying that it is cylindrical with very little
warping.

 In Fig.\ \ref{rotationplot} (lower panel), we show
results obtained in a second experiment on crystal A in which the $c$-axis was aligned at $\sim$15$^\circ$ from the
rotation axis and the magnetic field was rotated approximately within the basal plane. Two minima separated by
$60^\circ$ were observed, consistent with the hexagonal symmetry of MgB$_2$.   The minimum value of $F_3$ in the upper
panel agrees with that in the lower panel (after correcting for the 15$^\circ$ offset) showing that when
$\theta=90^\circ$, in the upper panel, $B$ lies along the [100] symmetry direction.

\begin{figure}
\includegraphics[width=8.0cm,keepaspectratio=true]{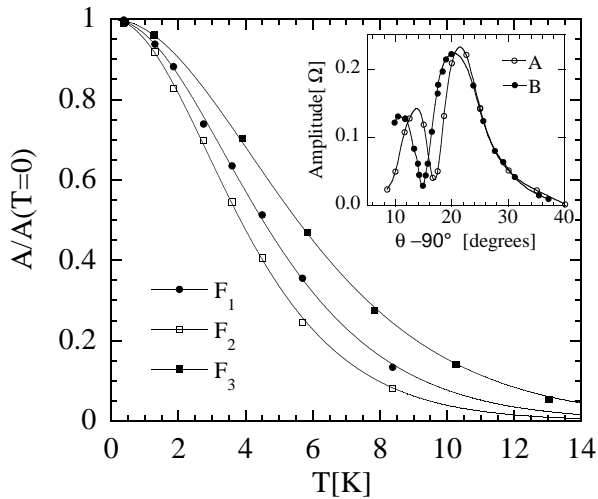}
 \caption{Dependence of dHvA amplitude on temperature.
  The amplitude was measured by fitting the sine function [Eq.\ (1)] to 1.5 oscillations in the range
 $17.70\rightarrow17.82$ T for $F_3$ and $16.6\rightarrow17.9$T for $F_1$ and $F_2$. The field was determined as the
 mean of $1/B$ in this window. The fit is to $R_T$ in Eq.\ (\protect\ref{lkexpression}).
  Inset: Dependence of amplitude on angle for $\theta$ near [100] for crystals A ($\circ$) and B ($\bullet$).
  Raw data are shown for A at 15~T and 1.35~K \protect\cite{footnote1} with data for B at 18~T
   and 0.36~K, normalized to the same peak height. }
 \label{massplot}
\end{figure}

Temperature dependent studies were made for all three frequencies from 1.35~K to 12~K (crystal A) and from 0.36~K to
12~K (crystal B) and analyzed in the standard way by fitting the data to the damping factor $R_T$ [Eq.\
(\ref{lkexpression})]. Fits for crystal B are shown in Fig.\ \ref{massplot}, and values of $m^*$ obtained in this way
for both crystals are given in \mbox{Table \ref{masstable}}. Measurements of $m^*$ were made with $B$ approximately
20$^\circ$ off a symmetry axis. The values in Table \ref{masstable} have been multiplied by $\cos(\theta)$ [or
$\sin(\theta)$] to correct for the usual angular dependence of the mass for a tubular band. This correction is less
than 8\%.

\begin{table}
\caption{Summary of results for both samples along with estimates of the frequencies and band masses from
band-structure calculations (c) \protect\cite{mazin02}.}
\begin{ruledtabular}
\begin{tabular}{r|r|r|r|r}
orbit&&$F_1^0$&$F_2^0$&$F_3^0$\\
\hline Freq. [T]&A&535 $\pm4$&1530 $\pm15$&2688 $\pm14$\\
&B&546 $\pm1$&1533 $\pm10$&2685 $\pm2$\\
&$c$&728&1756&2889\\
\hline Mass $m^*/m_e$
&A&0.53 $\pm$0.01&0.680$\pm$0.03&0.446$\pm$0.01\\
&B&0.553 $\pm$0.01&0.648 $\pm$0.01&0.441 $\pm$0.01\\
$m_B/m_e$&$c$&0.251&0.312&0.315\\
\hline $\tau$[ps], ($\ell$ [\AA]) &A&0.18 (500) &0.18 (660) &0.10 (750) \\&B&0.14 (380) &0.15 (580) &0.09 (680)
\\

\end{tabular}
\end{ruledtabular}
\label{masstable}
\end{table}

In the inset to Fig.\ \ref{massplot} we show the angular dependence of the dHvA amplitude near [100], for both
crystals.  There is a pronounced dip between 14$^\circ$ and 18$^\circ$.  The slight difference for the two crystals is
ascribed to small alignment errors about an axis perpendicular to the axis of rotation.  No such dip was observed for
$F_1$ and $F_2$. We believe the dip for $F_3$ is a `spin-zero', which is often observed in dHvA studies when the areas
of spin-up and spin-down extremal orbits differ by a half-integral number of Landau quanta, leading to destructive
interference in the oscillatory magnetization.  This dip was shown to persist to at least 4.2~K. Since $F_3$ comes from
a tubular surface, whose effective area $\propto 1/\cos\theta$, it is easily shown that such zeros occur at angles
$\theta_{sz}$ given by $g\frac{m_S}{m_e} = n\cos \theta_{sz}$, with $n = 1,3,5..$. As before, $\frac{m_S}{m_e}$
includes band effects and electron-electron enhancement but not electron-phonon enhancement of the standard type
\cite{shoenberg}. This is because the Pauli susceptibility of a metal can be enhanced by electron-electron
interactions, e.g., by the Stoner mechanism, but not by the electron-phonon interaction. Our data give
$\theta_{sz}=18\pm3^\circ$, and for $n=1$ and $g=2$ this gives $m_S = 0.476\pm0.007 m_e$.
\begin{figure}
\includegraphics[width=8.0cm,keepaspectratio=true]{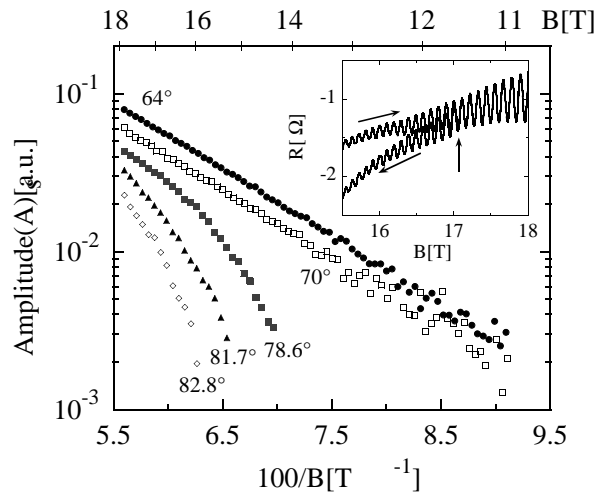}
 \caption{dHvA amplitude ($A_s$) at 0.36~K versus inverse field for several values of $\theta$.
 The inset shows up and down field sweeps at 81.7$^\circ$. The field sweep direction is indicated.
 The vertical arrow marks the point where the hysteresis vanishes.}
 \label{dingleplot}
\end{figure}

The scattering rates of the three orbits were extracted using two procedures. We either subtracted a smoothly varying
background and fitted the whole field sweep to Eq.\ (\ref{lkexpression}) (or more precisely, $\Gamma \propto B M_{osc}
\partial \ln F$/$\partial \theta $), or fitted Eq.\ (\ref{lkexpression}) to 1.5 oscillations in $1/B$ to give the
dHvA amplitude with the thermal damping factor removed ($A_s$), as a function of $B$. In both cases we used the values
of $m^*$ from Table \ref{masstable} in the $R_T$ factor and set $R_S$= 1. In field regions where the sample is clearly
in the normal state, the first method gives excellent fits to Eq.\ (\ref{lkexpression}) for all three frequencies.
Scattering times derived from fits for both crystals and all frequencies are shown in Table \ref{masstable}. Making the
approximation that all three frequencies arise from circular areas in $\mathbf{k}$-space, we can obtain $k_F$ for each
area and hence $v_F=\hbar k_F/m^*$, and the mean free path $\ell = v_F\tau$.  These are also shown in Table
\ref{masstable}.

The second procedure removes thermal damping effects  and reveals changes in the amplitude ($A_s$) versus $1/B$ plots
caused by the onset of superconductivity below $H_{c2}$. As shown in  Fig.\ \ref{dingleplot} at angles of 70$^\circ$ or
less, the usual exponential behavior is obtained, (with  slopes  $\propto$ 1/$\cos(90-\theta)$ as expected for a
tubular surface). However, for angles nearer to the plane, the Dingle plots become non-linear because of additional
damping caused by the growth of the superconducting gap, as has been observed in many superconductors (see for example
Ref.\ \cite{janssen98}). This observation proves that our dHvA signals arise from MgB$_2$ rather than any impurity
phase. The inset to Fig.\ \ref{dingleplot} shows hysteresis arising from superconductivity that roughly corresponds to
the onset of non-linearity in the Dingle plot. However, the slope of the Dingle plot for $\theta=81.7^\circ$ remains
high above 17~T, suggesting that the superconductivity has not been entirely suppressed in the reversible region.

Recently, detailed bandstructure calculations, including estimates of dHvA frequencies and masses, have been carried
out by three independent groups \cite{harima,mazin02,elgazzar02}. All calculations predict two warped cylinders along
the $c$ direction and two honeycomb networks in the basal plane. Comparison of the calculated frequencies with our
experimental data  shows that $F_1$ and $F_2$ arise from the smaller warped cylinder along the $c$-axis while  $F_3$
corresponds to the electron-like, in-plane tubular network whose median plane contains the A,H and L symmetry points.
The discrepancies with theory are less than 300 T (see Table \ref{masstable}) which is only 0.2 \% of the area of the
hexagonal Brillouin zone. Therefore, we can confidently compare our experimental values of $m^*$  with the calculated
values of $m_B$ to obtain the electron-phonon enhancement factors ($\lambda$).   We find $\lambda = (m^*/m_B-1)$= 1.20,
1.08 and 0.40 for $F_1$, $F_2$ and $F_3$ respectively, which compare favorably with the calculated values $\lambda$ =
1.25,1.25 and 0.47 (Refs.\ \cite{liu01,mazin02,choi02}).

The detailed angular dependence of $F_1$, $F_2$ and $F_3$ has been calculated by Harima \cite{harima}. We find this
also agrees with our data and supports the Yamaji analysis given above.  The calculated values of $m_B$ for $F_3$ also
allow us to estimate the Stoner enhancement of the susceptibility.  We find the enhancement $m_S/m_B-1 = 0.51$ on this
orbit, which is twice the calculated value of 0.26 \cite{mazin02}.

The remaining discrepancy with theory concerns the absence of other dHvA orbits in the present study, particularly
those arising from the second $c$-axis tube and the other in-plane network (band 4 in the notation of Ref.\
\cite{harima}).  In view of the discrepancies of 300 T mentioned above, it is possible that the smallest orbits ($<500$
T) are actually even smaller or absent.  The non-observation of the other orbits could well be due to the relatively
short mean-free-path of our crystals. Experiments to higher field and/or with purer crystals will clarify this point.

In summary, we have presented dHvA data for two single crystals of the new superconductor MgB$_2$ that are in excellent
agreement with each other and in good agreement with the most recent band structure calculations. The present work
provides direct evidence that the electron-phonon interaction is large on the inner $c$-axis cylinder and much smaller
on the in-plane honeycomb network. This supports microscopic theories of superconductivity that invoke two bands with
very different properties \cite{liu01,shulga01}.

JRC is grateful to D.\ E.\ Farrell for introducing him to the cantilever technique during a sabbatical year in
Cambridge in 1997 and to J.\ W.\ Loram and W.\ Y.\ Liang for their support. We thank S.\ Drechsler, H.\ Harima, S.\ M.\
Hayden, J.\ Kortus, I.\ Mazin, M.\ Springford, G.\ Santi, and  J.\ A.\ Wilson for helpful discussions. This work was
supported by the NEDO (Japan) as Collaborative Research and Development of Fundamental Technologies for
Superconductivity Applications and the EPSRC (U.K.). PJM gratefully acknowledges the support of the Royal Society
(London).


\begin{thebibliography}{99}
\bibitem{footnote1}  The piezolever sensitivity was $\simeq 10^{-10}$Nm/$ \Omega$ (low field measurements in the diamagnetic
state showed that this factor changed by less than $\pm$ 5 \% between 1.4~K and 10~K).  Typical noise levels were
$\pm 8$ m$\Omega$ ($\pm2\times 10^{-13}$Nm).
\end{thebibliography}
\end{document}